\documentclass{article}
\usepackage{spconf,amsmath,graphicx,multirow}
\usepackage{amsmath}
\usepackage{xcolor}

\title{Stack-and-Delay: A new CODEBOOK PATTERN FOR MUSIC
GENERATION}
\name{Gael Le Lan, Varun Nagaraja, Ernie Chang}
\name{\textit{David Kant, Zhaoheng Ni, 
Yangyang Shi, Forrest Iandola, Vikas Chandra}}

\name{%
\begin{tabular}{@{}c@{}}
Gael Le Lan \qquad 
Varun Nagaraja \qquad 
Ernie Chang \\
David Kant \qquad
Zhaoheng Ni \qquad
Yangyang Shi \qquad 
Forrest Iandola \qquad 
Vikas Chandra 
\end{tabular}}
\address{Meta AI}
\begin{document}
\maketitle
\begin{abstract}

In language modeling based music generation, a generated waveform is represented by a sequence of hierarchical token stacks that can be decoded either in an auto-regressive manner or in parallel, depending on the codebook patterns. 
In particular, flattening the codebooks represents the highest quality decoding strategy, while being notoriously slow. 
To this end, we propose a novel stack-and-delay style of decoding strategy to improve upon the \emph{flat} pattern decoding where generation speed is four times faster as opposed to vanilla flat decoding.  
This brings the inference time close to that of the \emph{delay} decoding strategy, and allows for faster inference on GPU for small batch sizes.
For the same inference efficiency budget as the \emph{delay} pattern, we show that the proposed approach performs better in objective evaluations, almost closing the gap with the \emph{flat} pattern in terms of quality. The results are corroborated by subjective evaluations which show that samples generated by the new model are slightly more often preferred to samples generated by the competing model given the same text prompts.

\end{abstract}
\section{Introduction}
\label{sec:intro}

The task of text-to-music generation has seen an increasing interest from the research community in the past year \cite{schneider2023mo,huang2023noise2music,lam2023efficient,copet2023simple,li2023jen,garcia2023vampnet}. 
This was enabled by the emergence of two competing architectures originating from the computer vision and natural language processing spaces, respectively: 
diffusion \cite{dhariwal2021diffusion,chang2022maskgit} and Transformer-based language models (LMs)~\cite{radford2019language,touvron2023llama}. 
The former method can be referred to as parallel decoding while the latter is usually auto-regressive.

The level of quality is getting closer to that of original songs, paving the road towards new commercial use cases such as personalized on-device music generation, where the batch size is typically small.
However those models often come with a quality trade off: the higher the quality, the slower the generation and vice versa \cite{lam2023efficient, garcia2023vampnet}.
During inference, the decoding strategy, hardware and model size influence the speed of the generation.
\cite{copet2023simple} recently proposed a single-stage auto-regressive Transformer decoder that models sequences of compressed discrete music representations (i.e. tokens compute by an audio compression model \cite{defossez2022high}). 
The authors explored several codebook patterns for the discrete tokens sequence modeling.
In particular, they showed that the best performing pattern relies on flattening the token stack (which will be referred to as the \emph{flat} pattern in the rest of the paper).
Indeed each piece of generated waveform is actually represented by not only one token but several, corresponding to the number $C$ of residual projections in the Residual Vector Quantizer (RVQ) \cite{zeghidour2021soundstream} module of the compression model.

Flattening the token stack comes with the cost of generating (and training) for a $C$ times longer sequence, which implies a significantly higher real-time-factor (RTF), making the model unusable in practice for interactive user experience. 
To overcome that issue, the proposed \emph{delay} pattern \cite{copet2023simple} was shown to be a good trade off between speed and quality.

In this paper we hypothesize that despite its efficiency, the \emph{delay} pattern could affect the model ability to generate high quality samples by design. Starting from the stronger but slower \emph{flat} pattern, we propose a new strategy called \emph{stack-delay} that is able to generate music as fast as the original \emph{delay} strategy, with significantly higher quality. The contributions of this paper are:
\begin{itemize}
\vspace{-4pt}
\item a new \emph{stack} codebook pattern that inherits the quality of \emph{flat} while being faster and memory efficient during inference by reducing the past key/value streaming cache footprint.
\vspace{-4pt}
\item a new \emph{stack-delay} pattern that:
\vspace{-4pt}
\begin{itemize}
\item benefits from the \emph{stack} pattern strengths while being as fast as the \emph{delay} pattern for generation.
\item produces higher quality music than \emph{delay}, shown by objective and subjective evaluations.
\end{itemize}
\vspace{-4pt}
\item an new decoding schedule that involves interleaving decoded positions that prevents the model from decoding adjacent positions until they have enough context.
\end{itemize}
\vspace{-4pt}

\begin{figure*}[tb]
\centering
\includegraphics[width=1.00\textwidth]{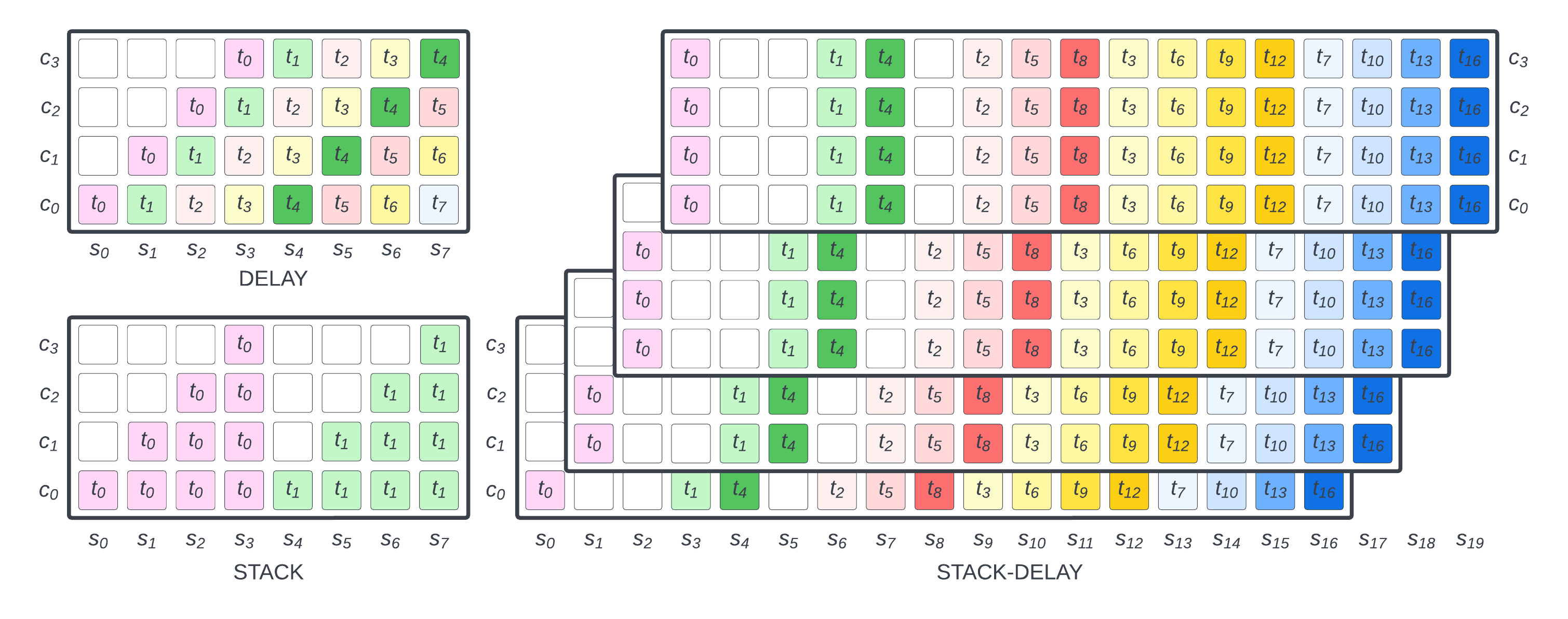}

\caption{Comparison of the proposed \emph{stack-delay} pattern (right) with the \emph{delay} (top left) and \emph{stack} (bottom left). Under the \emph{stack-delay} pattern the tokens are generated in a multi-stream fashion, in parallel. Time steps are decoded in a permuted manner. Only key/value embeddings from the top-level stream are stored in long-term streaming cache, which makes inference as efficient as \emph{delay} while retaining better quality from \emph{stack} pattern.}
\label{fig:interleaving}
\end{figure*}

\section{Stack-delay codebook pattern}
\label{sec:paradelay}

\subsection{Music generation}
Given a text description, a sequence of text embeddings computed by the T5 encoder \cite{raffel2020exploring} serves as the conditioning signal for a Transformer decoder model (using cross attention). The model generates a sequence of EnCodec \cite{defossez2022high} token stacks $\{c_{it}\}_{i=0}^{C-1}$ that are CNN-decoded into an audio waveform. $i$ represents the token level while $t$ represents the time step in the generated sequence.

In this paper we only consider the auto-regressive Transformer decoder architecture \cite{radford2019language} that emits a probability distribution over the token space that is conditioned on the previously generated tokens (causal self attention in the Transformer decoder). During inference, the past self attention keys and values are stored in a streaming cache to optimize the generation time.
Depending on the tokenizer framerate $f$ (e.g. $f=50Hz$), the duration of audio to generate $d$ and the size of the token stack $C$ (e.g. $C=4$), the model has to generate $f \times C \times d$ tokens in a given amount of decoding steps that depend on the token codebook pattern and decoding schedule. The decoding schedule can be formalized as a function $\mathcal{G}(i,t)$ defining the decoding step for each $c_{it}$.

\subsection{Codebook patterns}

Contrary to the text domain, a segment of audio is not represented by a single token but by a stack of hierarchical tokens computed by quantizing \cite{zeghidour2021soundstream} the latent embedding of a CNN auto-encoder \cite{defossez2022high}. This usually means the lower the token in the stack, the more information it carries. To address the issue of predicting tokens in a hierarchical manner, several codebook interleaving patterns have been explored \cite{wang2023neural, copet2023simple, borsos2023soundstorm}, with the common idea to decode the lowest level token first then handle the higher levels in further decoding steps, which is the case for both auto-regressive (AR) \cite{copet2023simple} and non auto-regressive (NAR) \cite{garcia2023vampnet} decoding architectures. Namely the decoding schedule is constrained such that:
\begin{equation}
\mathcal{G}(0,t) < \mathcal{G}(i,t), \forall i \in [1, C[
\end{equation}

\subsubsection{\emph{Delay}}

Regarding music generation, the \emph{delay} interleaving pattern (presented on the top left part of Figure \ref{fig:interleaving}) was shown to be a good compromise between quality and AR decoding step count. Under the \emph{delay} pattern, the $C$ codebook levels are predicted in parallel but with a shift of in the decoded time step. Namely $\mathcal{G}(i,t) = t + i$. This means that each subsequent time step in the sequence starts to be decoded with only partial knowledge of the previous adjacent time step. For example, the prediction of $c_{0t_1}$ in decoding step $s_1$ in the Figure is only conditioned on $c_{0t_0}$, previously decoded in $s_0$, but not on higher levels $\{c_{i}\}_{i=0}^{C-1}$ of time step $t_0$.

\subsubsection{\emph{Stack}}

\cite{copet2023simple} showed that to obtain the highest music quality, flattening the codebooks performed the best, at the expense of $C$ times more decoding steps.
\begin{equation}
    \mathcal{G}(i, t) = C \times t + i < C \times T
\end{equation}
This can be easily explained by the fact that subsequent decoded time steps benefit from the full context of the preceding ones. In such case the prediction of $c_{0t+1}$ is effectively conditioned on $c_{[0,C-1][0,t]}$.
The context length is $C$ times bigger than \emph{delay} since the at most $C \times T$ past Transformer self attention key/value representations are stored in the streaming cache during inference.
To reduce the cache size we adapt the \emph{flat} pattern by retaining and stacking the lower level tokens throughout the decoding process, as shown in Figure \ref{fig:interleaving}. Once a full stack has been decoded for a given time step, partial stacks can be erased from the streaming cache as the full stack contains all the required information. This way the maximum cache length is only of $C + T$ instead of $C \times T$. The \emph{stack} pattern requires a customized attention mask during training that simulates the inference dynamic caching behavior. However it still requires $C$ times more decoding steps than \emph{delay}.

\subsubsection{\emph{Stack-delay}}

To compensate for the increased decoding step count (i.e. inference time) of the \emph{stack} pattern, we propose to introduce $C$ parallel decoding streams in what we call the \emph{stack-delay} pattern, illustrated in the right part of Figure \ref{fig:interleaving}.
Having $C$ parallel streams decoding a $C$ times longer sequence means that overall the total number of decoding steps is the same as for the \emph{delay} pattern (i.e. $T$). The main difference with \emph{delay} is that we no longer stack tokens from different time steps but always from the same time step. This also allows positional encoding to encode not only the decoded time step but also the decoded token level, hence hinting the model about which time step and level is about to be decoded. We hope this will improve the overall model performance for the same inference efficiency budget as \emph{delay}, due to the use of parallel-optimized compute hardware. We report the decoding step count and maximum context length in Table \ref{tab:steps} for each pattern.

\begin{table}[bt]
\centering
\footnotesize
\begin{tabular}{l|c|c}
\hline
pattern & decoding steps & context length \\
\hline
\hline
\emph{delay} & $T$ & $T$ \\
\emph{flat} & $T \times C$ & $T \times C$ \\
\emph{stack} & $T \times C$ & $T + C$ \\
\emph{stack-delay} & $T$ & $T$ \\
\hline
\end{tabular}  
\caption{Required decoding step count and maximum context length of the streaming cache during inference, as a function of the sequence length to generate $T=d \times f$ and the number of codebook levels $C$.}
\label{tab:steps}
\end{table}

\subsubsection{Timesteps interleaving}
\label{subsec:schedules}

Finally, we introduce time steps permutation in the decoding schedule: the decoding remains auto-regressive but the model is trained to predict the token sequence in a time step-permuted order. This aims to offer more context for adjacent time steps decoding.
An example of such interleaving pattern is shown on the right part of Figure \ref{fig:interleaving}, which corresponds to the decoding schedule defined in equation \ref{eq:decoding} with $k = 3$. According to the equation, the \emph{delay} pattern decoding schedule corresponds to the case where $k=1$.
\begin{equation}
\label{eq:decoding}
\mathcal{G}(i,t) = t + (t\mod (k+1)) \times (k-1) + i
\end{equation}

\section{Experimental setup}
\label{sec:typestyle}

Most of the experimental setup follows that of MusicGen \cite{copet2023simple}, we refer the readers to it for more details.

\subsection{Model}

The tokenizer is an EnCodec model \cite{defossez2022high}, made of CNN autoencoder and Residual Vector Quantization module applied to the latent representation of waveforms. The RVQ module is made of $C=4$ quantizers, each with a codebook size of 2048. It encodes 32 kHz monophonic audio into a stack of 4 tokens every 20ms (50 Hz framerate).

The Transformer decoder is made of 300M parameters, implemented with a customized version of \emph{audiocraft}\footnote{https://github.com/facebookresearch/audiocraft}. It uses Pytorch 2.0\footnote{https://pytorch.org/} flash attention for faster training and generation with optimized memory footprint. The model is trained on 30-seconds random crops of the full track. The models are trained for 200 epochs (400k steps) with the AdamW optimizer, a batch size of 192, $\beta_1$ = 0.9, $\beta_2$ = 0.95, a decoupled weight decay of 0.1 and no gradient clipping. A cosine learning rate schedule with a warmup of 4000 steps is used at the beginning of training. Models are trained with an exponential moving average with 0.99 decay. Training uses \textit{fp16} mixed precision and distributed data parallelism on 24 A100 GPUs.

\subsection{Generation}

At each decoding step the Transformer decoder emits a probability distribution over the token space for time steps and levels to decode according to the decoding schedule. Tokens are sampled from the distribution with top-$k$ nucleus sampling with $k=250$ tokens and a temperature of 1.0. We apply classifier-free guidance \cite{kreuk2022audiogen} when sampling from the model’s logits, with a guidance scale of 3.0.

The baseline model uses the \emph{delay} codebook pattern from \cite{copet2023simple}. This translates 30 seconds of audio into $T=500$ auto-regressive steps. For text conditioning, we use the T5 \cite{raffel2020exploring} text encoder. During training we drop the text condition with a probability of 0.1. We experiment with \emph{flat}, \emph{stack} and \emph{stack-delay} codebook patterns.

\subsection{Data}
We train our models on 20K hours of licensed music: an internal dataset of 10K high-quality music tracks and the ShutterStock and Pond5 music data collections\footnote{www.shutterstock.com/music and www.pond5.com} with respectively 25K and 365K instrument-only recordings. All recordings are sampled at 32 kHz and come with a textual description. The models are evaluated on an in-domain split different from that of \cite{copet2023simple} and on the MusicCaps dataset \cite{agostinelli2023musiclm}.

\subsection{Evaluation}

The different models are evaluated through a set of generated samples from a list of evaluation text-prompts. For objective evaluation we compute Frechet Audio Distance (FAD) using VGG classifier \cite{hershey2017cnn}, Kullback–Leibler divergence (KLD) using PaSST model \cite{koutini2021efficient}, and CLAP similarity score \cite{elizalde2023clap}. For subjective evaluation we run a blind pairwise comparison test where we present the evaluator two samples generated by two models but using the same text prompt, for a list of 20 text prompts. The human evaluators are asked to select the preferred sample from each pair based on perceived quality. Finally we report the RTF computed on A100 GPU when generating one sample (effective batch size of 2 from the model perspective due to classifier free guidance).

\section{Results}
\label{sec:results}

\begin{table}[bt]
\centering
\begin{tabular}{l|r|c|c|c|c}
\multirow{2}{*}{pattern} & \multicolumn{3}{|c|}{in-domain} & MusicCaps & RTF \\
&FAD & KLD & CLAP& FAD & (A100) \\
\hline
\hline
\emph{delay} & 0.69 & 0.48 & 0.36 & 4.91 & 1.07 \\
\emph{flat} & 0.42 & 0.47 & 0.37 & 5.25 & 4.69 \\
\emph{stack} & \textbf{0.38} & 0.48 & 0.37 & 5.16 & 4.77 \\
\emph{stack-delay} & 0.48 & 0.48 & 0.37 & \textbf{4.88} & 1.13\\
\end{tabular}%
\caption{Quality/efficiency trade off of the proposed token sequence patterns for 30 seconds generated tracks.}
\label{tab:results}
\end{table}

\begin{table}[bt]
\centering
\begin{tabular}{l|r|c|c}
decoding schedule $\mathcal{G}(i,t)$ & FAD & KLD & CLAP \\
\hline
\hline
$t + i$ (\emph{delay}) & 0.45 & 0.50 & 0.38 \\
\hline
\hline
$t + i$ (\emph{stack-delay}) & 0.43 & 0.51 & 0.37 \\
$t + (t\mod 3) \times 1 + i$ & 0.42 & 0.50 & 0.37 \\
$t + (t\mod 4) \times 2 + i$ & 0.36 & 0.51 & 0.38 \\
$t + (t\mod 5) \times 3 + i$ & \textbf{0.34} & 0.52 & 0.38 \\
\end{tabular}
\caption{Ablation study on the effect of permuting timesteps in the decoding schedule of the \emph{stack-delay} pattern, for 10s samples on the in-domain evaluation dataset.}
\label{tab:timesteps}
\end{table}

\subsection{Baselines - \emph{flat} and \emph{delay} patterns}

We consider two baselines: \emph{flat}, which is known to produce the highest quality audio although requiring much more compute than \emph{delay}, and \emph{delay}, a good compromise between speed and performance, achieving a RTF close to 1, potentially unlocking streaming scenarios. \emph{flat} achieves an in-domain FAD of 0.42, 39\% lower than \emph{delay}, while KLD and CLAP remain close. Despite its higher quality the RTF is above 4.

\subsection{\emph{Stack} pattern}

We first investigate the \emph{stack} pattern as a replacement for the (so far) state-of-the-art \emph{flat}. Our results indicate that it is competitive with \emph{flat}, even outperforming its FAD score with 0.38, with a similar RTF. The better FAD score indicates that the shorter required context length for generation might have a positive effect on music quality for long samples generations.

\subsection{\emph{Stack-delay} pattern}

When considering the \emph{stack-delay} pattern, our results indicate that it outperforms \emph{delay} with a FAD of 0.48, although not as low at \emph{stack}, but much more efficient with almost the same RTF as \emph{delay}, unlocking potential real time streaming scenarios with better quality than the baseline.
For subjective evaluation we only compare the \emph{stack-delay} and \emph{delay} versions. Our results indicate that samples generated by the \emph{stack-delay} are preferred 51.3\% of the time compared with \emph{delay}. Such a small difference is to be expected given the small scale of our subjective evaluation.

\subsection{Ablation - permuting decoded time steps}

Finally, we look into the interleaved time steps decoding schedules defined in section \ref{subsec:schedules}. The ablation results are presented in Table \ref{tab:timesteps} that compares four different schedules applied with the \emph{stack-delay} pattern, and also including the \emph{delay} baseline.

The table shows that the higher the decoding step count separating adjacent positions, the lower the FAD, with KLD and CLAP scores in a similar range. This shows the benefit of permuting the time steps in the \emph{stack-delay} pattern. Without permutation (i.e. following the same ascending time steps schedule as \emph{delay}), the \emph{stack-delay} pattern only achieves marginal improvement. We also tried applying the \emph{delay} pattern with the same permuted schedules and the performance was only on par with the baseline, which means that the combination of the proposed pattern and permuted decoding schedule is essential for better performance.

\section{Conclusion}
\label{sec:print}

We introduce a new codebook pattern that relies on stacking the discrete music tokens, delaying/shifting the decoding of subsequent levels, and permuting the order of time steps to decode in the decoding schedule. The combination of the three outperforms the \emph{delay} baseline quality-wise with a in-domain FAD reduction of 45\% for the same inference efficiency budget, due to parallel decoding that compensates for an increased sequence length. We also show that stacking the tokens should be preferred to flattening them best when the highest quality is a priority.
Finally the ablation study shows that time step permutation is key to achieve optimal performance, indicating that decoding of adjacent positions with only partial knowledge of previous time steps probably affects the performance of the \emph{delay} pattern. Overall we hope our findings can help design better non-autoregressive decoding strategies in the future.

\vfill\pagebreak

\bibliographystyle{IEEEbib}
\bibliography{main}

\end{document}